# A Nanomembrane-Based Bandgap-Tunable Germanium Microdisk Using Lithographically-Customizable Biaxial Strain for Silicon-Compatible Optoelectronics

*David S. Sukhdeo[†,*,#], Donguk Nam[†,#], Ju-Hyung Kang[‡],*

*Mark L. Brongersma[‡] and Krishna C. Saraswat[†]*

[†]Department of Electrical Engineering, Stanford University, Stanford, CA 94305, USA,

[‡]Department of Materials Science and Engineering, Stanford University, Stanford, CA 94305, USA

**ABSTRACT**: Strain engineering has proven to be vital for germanium-based photonics, in particular light emission. However, applying a large permanent biaxial strain to germanium has been a challenge. We present a simple, CMOS-compatible technique to conveniently induce a large, spatially homogenous strain in microdisks patterned within ultrathin germanium nanomembranes. Our technique works by concentrating and amplifying a pre-existing small strain into the microdisk region. Biaxial strains as large as 1.11% are observed by Raman spectroscopy and are further confirmed by photoluminescence measurements, which show enhanced and redshifted light emission from the strained microdisks. Our technique allows the amount of biaxial strain to be customized lithographically, allowing the bandgaps of different microdisks to be independently tuned in a single mask process. Our theoretical calculations show that this platform can deliver substantial performance improvements, including a >200x reduction in the lasing threshold, to biaxially strained germanium lasers for silicon-compatible optical interconnects.

With electrical interconnects emerging as a severe performance bottleneck in silicon (Si) complementary metal-oxide-semiconductor (CMOS) devices, optical interconnects have become a leading contender for future CMOS technology[1,2]. However, Si's indirect bandgap limits its use in optoelectronics[2,3] and there remain substantial manufacturing and cost concerns with hybrid Si/III-V approaches. It would therefore be advantageous if an optical link on Si can be realized using only group IV materials[3]. As such, germanium (Ge) has garnered much attention[3–6] recently due to advances in Ge-on-Si heteroepitaxy[7] and because of Ge's ability to perform useful optoelectronic functions[8]. Although Ge is nominally an indirect bandgap semiconductor, it also has a direct bandgap of 0.800 eV which is only 133 meV larger than its indirect bandgap of 0.667 eV[9]. This difference is small enough that researchers have successfully realized efficient Ge-on-Si photodetectors[10–12] and modulators[13]. An electrically-pumped Ge-on-Si laser has also been demonstrated, but with an enormous threshold of ~300 kA/cm$^2$ which necessitates drastic improvements[14,15]. Two approaches, n-type doping and tensile strain, have been proposed to remedy the situation[16] and strain is a particularly promising route to an efficient, low-threshold Ge-on-Si laser[17]. Tensile strain improves the performance of Ge lasers by narrowing the direct bandgap relative to the indirect bandgap, with 4.6% uniaxial strain or 1.7% biaxial strain yielding a direct bandgap as calculated in Figure 1 using Ge's deformation potentials[9].

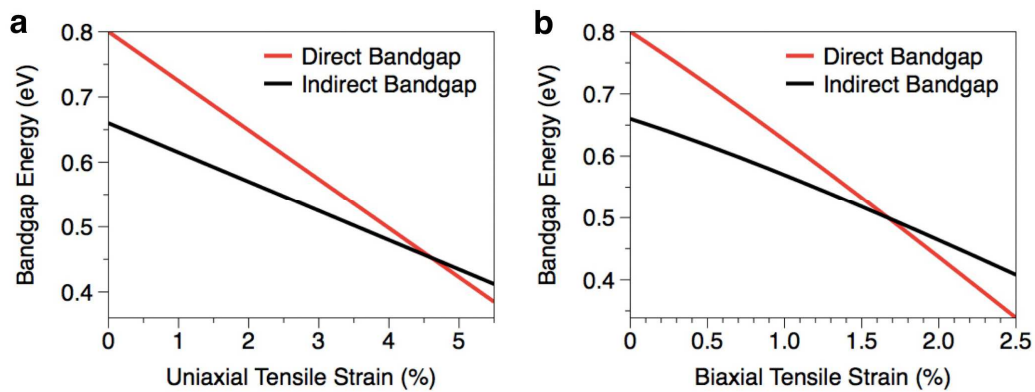

**Figure 1.** Theoretical calculations of bandgap changes as a function of strain. (a) Bandgaps vs. uniaxial strain, showing the direct bandgap cross-over at 4.6%. (b) Bandgaps vs. biaxial strain, showing the direct bandgap cross-over at 1.7%.

Tremendous advances have been made in realizing large uniaxial strains in Ge using nanowire[18] and micro-bridge structures[19], including a maximum reported strain of 5.7% which represents direct bandgap Ge-on-Si[20]. Similar advances for biaxially strained Ge-on-Si, however, have been lacking even though biaxial strain is best suited to the radial symmetry of microdisk and microgear resonators which combine high Q factors and compact form factors. While biaxially strained direct bandgap Ge has been demonstrated using pseudomorphic growth on lattice-mismatched GaAs/InGaAs substrates[21] or by temporarily applying gas pressure to a suspended Ge nanomembrane[22], neither approach produces a permanently-sustained strain in Ge integrated on a Si platform. A permanent biaxial strain of 1.1% was achieved in a freestanding Ge membrane supported on a Si substrate with a tungsten stressor[23], but the tungsten metal adjacent to the Ge and the out-of-plane deflection would compromise any optical cavity design. Lastly, a truly CMOS-compatible structure has been shown by depositing a stressed silicon nitride layer on a Ge stripe to induce an "equivalent biaxial strain" of up to 0.9%[24], however this strain was not truly biaxial and, critically, was very inhomogeneous in the vertical direction. There is therefore a need for a CMOS-compatible structure that induces large homogeneous biaxial tensile strains in Ge-on-Si that can truly be an effective platform for Ge lasers. Here we present such a structure and report experimentally measured biaxial tensile strains up to 1.11%. This strain can be conveniently customized from one device to another by lithographically modifying the dimensions of each structure, thereby allowing multiple strains – and therefore multiple bandgaps – to be realized across a single wafer in a simple one-mask process.

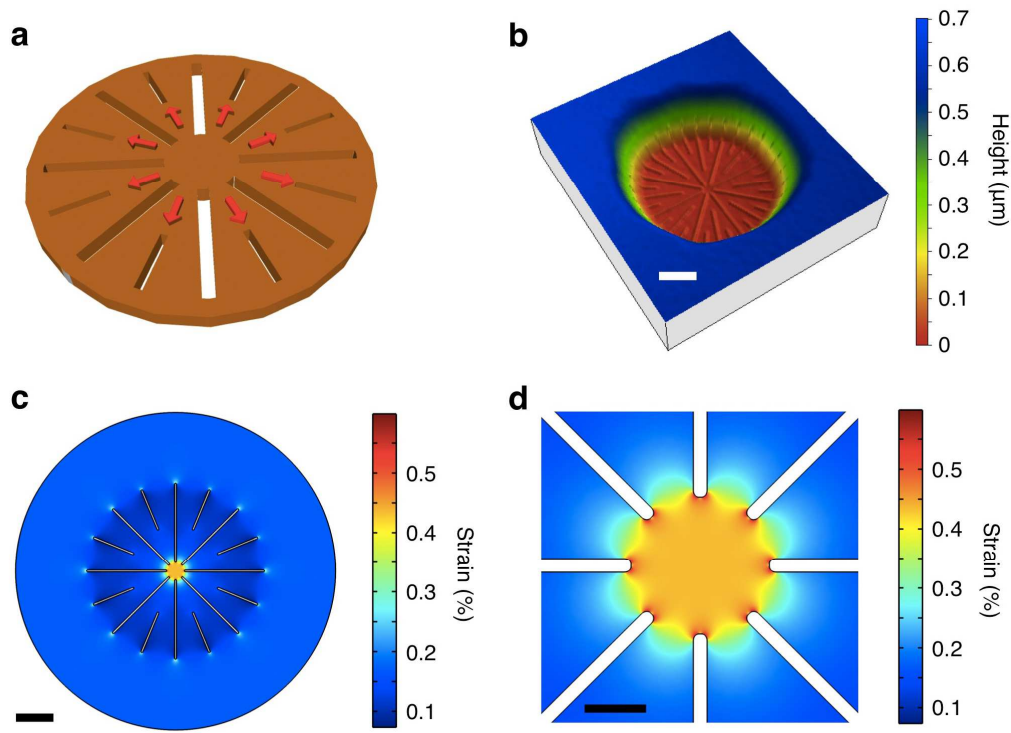

**Figure 2.** Overview of the biaxial strain structure. (a) Simplified schematic of the structure. (b) Optical surface profile of a fabricated structure. Note that the etched lines appear raised due to limitations of the optical interferometer setup; in reality these etch features are ~300 nm deep valleys in the profile instead of ~40 nm high raised features. Scale bar, 20 µm. (c,d) Biaxial strain distribution in the disk structures computed by FEM, shown at various zoom levels. Scale bar is (c) 10 µm, (d) 2 µm.

In order to achieve a large Ge volume under a tensile strain that is both large in magnitude and spatially homogenous, we have created a structure which concentrates a pre-existing strain biaxially using a special geometry, in analogy with previous work which used a micro-bridge geomtery[19,20,25,26] to concentrate a pre-existing strain uniaxially. Our structure consists of an ultrathin Ge disk with several etch slits in a radially symmetric pattern as shown in a simplified schematic (Figure 2a) which helps visualize the stress concentration process intuitively. Using electron-beam lithography, the top Ge layer of our Si/SiO$_2$/Ge substrate is patterned and etched into a disk with many slits as shown in the fabrication process flow (Figure S1 in the Supporting Information) During the final fabrication step, the oxide from the material stack is removed and the small pre-existing tensile stress redistributes and concentrates in the microdisk at the center of each structure. The small microdisks with diameters of 5.0-7.5 µm therefore become very highly strained, while most of the remaining Ge in the structure relaxes somewhat to compensate. Each entire structure was permanently adhered to the underlying silicon substrate due to stiction after release[27] as evidenced in the surface profile of the fabricated structure (Figure 2b). The strain in the inner microdisk is purely a function of the initial stress in the Ge and the ratio of the inner microdisk diameter to the total structure diameter. The initial stress in the Ge is usually ~0.2% which arises from the mismatch in the thermal expansion coefficient between Ge and Si during the epitaxial growth of Ge on Si[28]. By varying the ratio of the total diameter to the inner microdisk diameter, the strain in the microdisk can be varied lithographically from device to device. By varying this ratio, it is also possible to make the strain arbitrarily large, limited only by material fracture of the Ge. Importantly, only the eight etch slits touching the inner microdisk are vital to understanding the strain distribution. The additional etch slits in Figure 2b (i.e. all slits other than the eight touching the inner microdisk) were present only to facilitate the lateral etching of the underlying oxide by exposing more of this oxide as described in the Methods section in the Supporting Information. The additional slits could just as well have been a series of holes or any other shape that would reduce the maximum distance between exposed areas in order to facilitate lateral etching. The only effect of the additional slits on the strain distribution is to slightly reduce the strain in the inner microdisk, as shown in Figure S2 in the Supporting Information, and this effect can easily be compensated by slightly increasing the structure's total diameter.

We performed finite element method (FEM) modeling of the strain distribution in the structure using COMSOL, shown in Figure 2c for the case of a 5 µm inner microdisk diameter and a 50 µm outer (total) diameter, with 20 µm of lateral under-etching. Further FEM studies confirmed that the strain can indeed be made arbitrarily large by increasing the total (outer) diameter. The biaxial strain at the center of the inner microdisk (inner 5 µm diameter region) of Figure 2d is 0.435%, and remains constant to the third decimal place over an area of >10 µm$^2$. This represents a negligible variation over a substantial area, indicating an extremely homogeneous strain. This variation remains negligible according to our FEM simulations even as the total diameter, and hence the strain in the inner microdisk, is increased. Although already negligible, this variation can also be further reduced by increasing the number of etch slits touching the inner microdisk, as shown by additional FEM simulations in the Supporting Information. These FEM simulations also point to the expected failure mechanism for very highly strained disks: corner stresses near the inner edges of the main eight etch slits. This is plainly seen in the zoomed-in FEM simulations of Figure 2d. For this particular simulation, the biaxial strain, to be the average of $\varepsilon_x$ and $\varepsilon_y$ for the purposes of FEM simulations, reaches as high as 0.60%, even though the center of the inner microdisk is under only 0.435% biaxial strain, indicating a fracture risk near the corners. Thus, for devices with dimensions yielding larger strains than Ge can tolerate, we expect failure of these devices to occur by material fractures originating near these corner regions.

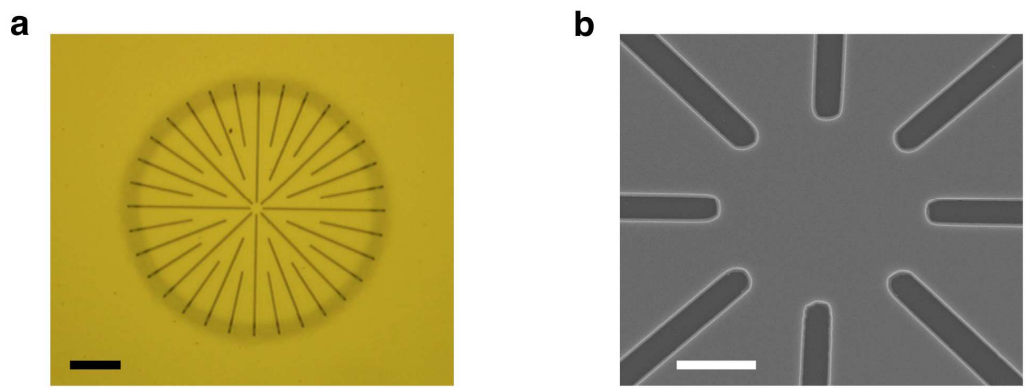

**Figure 3.** Micrographs of successfully fabricated structures. (a) Optical micrograph. Scale bar, 20 µm. (b) Scanning electron micrograph. Scale bar, 2 µm.

Structures were successfully fabricated with an inner microdisk diameter of 5 µm and with outer (total) diameters ranging from 30–130 µm, as shown by optical and scanning electron micrographs in Figure 3. A few devices were also successfully fabricated with 7.5 µm inner microdisk diameters for use in photoluminescence (PL) measurements. Several structures with larger outer diameters were attempted, however these devices failed due to material fracture as shown in Figure S4 in the Supporting Information, with fracture lines typically emanating from the corner regions as expected from the FEM simulations of Figure 2d. A study of several failed devices also revealed a strong tendency of the fracture lines to be aligned with the [110] cleavage planes, i.e. the standard cleavage planes for face-center cubic crystalline materials such as Ge and Si[29].

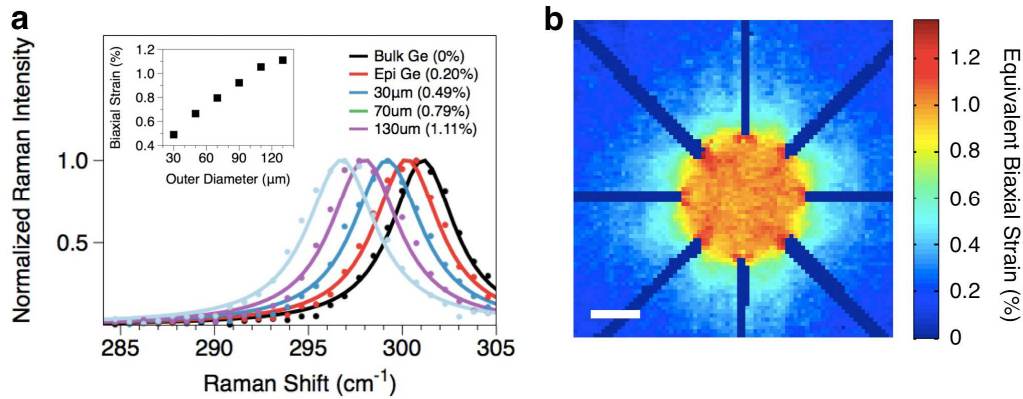

**Figure 4.** Raman Analysis. (a) Raman spectra for various structure diameters. Inset: observed strain vs. outer diameter. (b) Biaxial strain distribution in the structure, measured by a Raman area scan. Scale bar, 2 μm.

The strain in successfully fabricated structures was then measured by Raman spectroscopy, where a strain-shift coefficient of 390 cm$^{-1}$ was used following the method of Ref [30]. According to the observed Raman spectra of Figure 4a, the observed biaxial strains in the structures' inner microdisks ranged from 0.2% to 1.11%; the lower bound of 0.2% strain represents the residual Ge strain in the absence of patterning. The relationship between the observed strain and the outer diameter is shown in the inset of Figure 4a for a series of devices fabricated side-by-side in the same run. From Figure 4a it is clear that the measured strain has not saturated even for our largest unbroken sample (130 μm outer diameter), and FEM simulations predict no hard limit on the achievable strain if fracturing is ignored. Thus if the material fractures can be eliminated, perhaps by reducing the initial defect density in the Ge or by alleviating the corner stresses from which the fractures originate, even larger biaxial strains may be within reach with larger dimensions.

Another important advantage of our structure is that the biaxial strain is very homogenous over a large area, as suggested by our FEM modeling in Figure 2d. To confirm this homogeneity we have performed a Raman area scan and converted the measured Raman shifts to an "equivalent biaxial strain" by following the example of Ref [30]. As shown in Figure 4b, the observed strain is nearly constant over a large area in the inner microdisk, indicating an excellent strain uniformity except for some sharp increases near the corner regions. While the use of "equivalent biaxial strain" here is imperfect for very strongly non-biaxial strains in the corner regions, this is a non-issue over most of the central region where the strain is observed to be quite homogenous except for small variations which we ascribe to noise in the Raman measurements. This confirms that our structure achieves a uniform strain over a large area in practice as well as in theory.

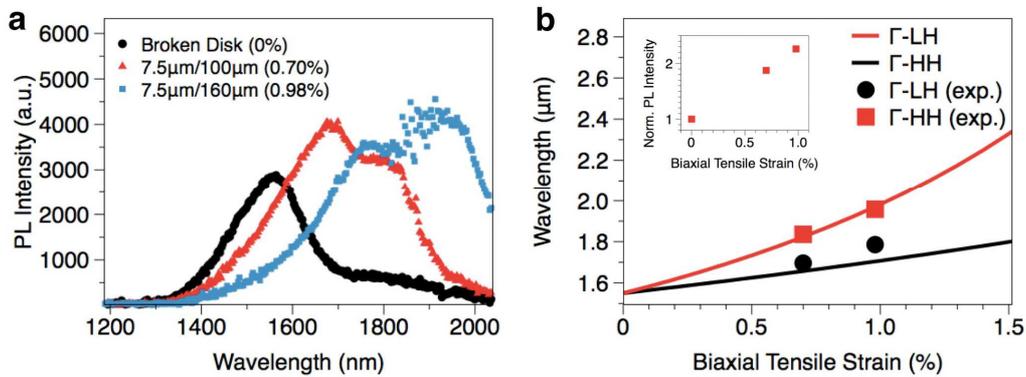

**Figure 5.** Photoluminescence. (a) Photoluminescence (PL) spectra at various strain levels. (b) PL wavelength vs. strain. Inset: PL intensity vs. strain.

Finally, the biaxially strained structures were characterized by PL measurements, as shown in Figure 5a, taken at the center of the inner microdisks. The structures characterized by PL had a relatively large inner microdisk diameter of 7.5 µm to accommodate the finite spot size of the excitation laser, and the use of substrate-adhered Ge structures in this work precludes any significant heating effects from the 12 mW laser excitation[20]. Biaxial strain is well understood to enhance Ge luminescence by increasing the fraction of electrons in the direct conduction valley[5,17,23,31,32], and we observe this phenomenon in our PL measurements. As shown explicitly in the inset of Figure 5b, the integrated intensity of the PL emission from our Ge structures increases by a factor of ~2.3x as the strain increases from zero to 0.98%. This is somewhat smaller than the ~20x enhancement in Gamma valley occupancy expected from Ge's deformation potentials[9], discussed in the next section, however this discrepancy can be explained by the fact that valence band splitting and polarization selection rules favor in-plane emission over out-of-plane emission as the biaxial strain increases[22]. Moreover, the presence of a large and uniform strain in the disks is unambiguously confirmed by analyzing how the wavelength of the PL emission changes with strain, and by confirming that the redshifts follow the energy separations between the Gamma (Γ) valley and the two separate valence bands, the heavy hole (HH) and light hole (LH) bands, as has been observed in previous works[33,34]. As shown in Figure 5b, employing larger strains does indeed redshift the PL emission in accordance with the narrowing of the Γ–HH and Γ–LH bandgaps predicted by theory.

While the achievement of a 1.11% biaxial strain is meaningful in its own right by virtue of how close this is to achieving a direct bandgap in Ge, particularly given that the strain is permanently sustained and maintains CMOS compatibility, it is instructive to theoretically investigate how this biaxial strain will affect Ge's optical properties. Using a tight-binding formalism with band edges based on the deformation potentials of Ref [9], the percentage of electrons residing in the direct conduction valley was calculated in Figure 6a for various n-type doping levels at room temperature. At zero doping, the observed strain of 1.11% in the 5 μm inner microdisk represents a ~15x enhancement of the direct conduction valley occupancy compared to 0.2% strain, the residual strain in epitaxial Ge[16,35] and the strain used in all demonstrated Ge lasers to date[14,36]. Comparing 0% strain and 0.98% strain, the bounds of our PL characterization, the predicted enhancement is ~20x for undoped Ge. For heavily n-doped Ge the predicted enhancements from strain are smaller but still substantial; an extended discussion of the interaction between strain and doping is provided in the Supporting Information. Interestingly, we observe that even for biaxial strains >1.7% which yield a direct bandgap, a majority of electrons will continue to reside in the indirect L conduction valleys due to their larger density of states (DOS) unless a strain well in excess of ~2.5% is employed. Although it is important to induce larger strains to improve the Ge's optical properties, it would be foolhardy to focus singularly on reaching a direct bandgap since achieving a direct bandgap results in no abrupt change in performance of any Ge device, with the exception of extremely low temperature applications[20].

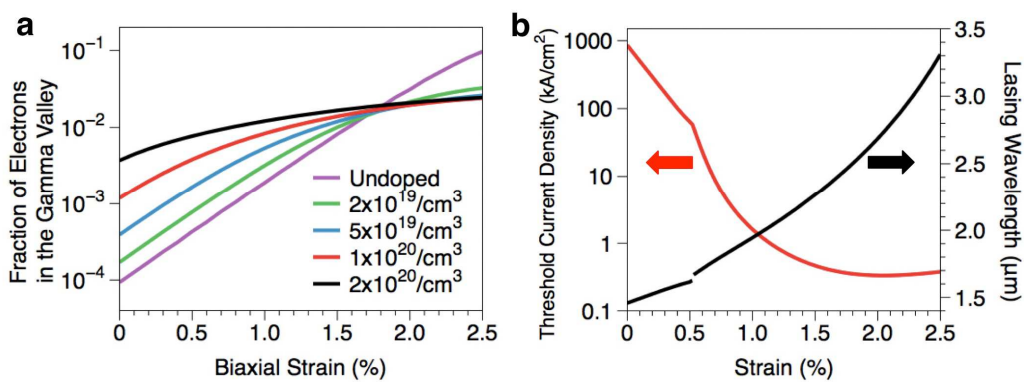

**Figure 6.** Theoretical modeling. (a) Fraction of electrons in the direct (Γ) conduction valley vs. biaxial strain at various n-type doping levels. (b) Threshold current density and lasing wavelength vs. biaxial strain, assuming a 300 nm thick Ge active region and a lossless optical cavity.

Several theoretical studies[16,17,37–39] have shown that biaxial strain will enable an efficient low-threshold Ge-on-Si laser; the direct conduction valley can dominate recombination processes even with only a small fraction of electrons due to rapid inter-valley scattering[40]. While these theoretical studies vary slightly in their predictions due to methodological differences, they all conclude that applying biaxial strain on the order of 1.11% will result in substantially larger optical gain and threshold reductions of several orders of magnitude. In particular, our theoretical model shows that beyond approximately 0.5% strain there will be an abrupt decrease in the lasing threshold due to valence band splitting effects[17] which simultaneously cause a discontinuity in the wavelength. This behavior is shown explicitly in Figure 6b assuming $4\times10^{19}$ cm$^{-3}$ n-type doping in accordance with the parameters of the experimentally-demonstrated Ge laser[14]. This means that even though 1.11% biaxial strain does not represent a direct bandgap in Ge, it has passed the critical value of ~0.5% strain where outsized reductions in the lasing threshold become possible. In particular, this 1.11% biaxial tensile strain is expected to enable a >200x reduction in the lasing threshold compared to the 0.2% biaxial strain that was used in the state-of-the-art electrically-pumped Ge laser[14].

In summary, we have experimentally achieved large biaxial strains of up to 1.11% in microdisks fabricated in Ge nanomembranes. Unlike previous works on highly-strained Ge[21–24], our strain satisfies all the conditions of being permanently-sustained, vertically and laterally homogenous over arbitrarily large volumes, and integrated directly on a silicon substrate while preserving full CMOS-compatibility. Our process has the further advantage of reduced cost and simplicity since it involves only one lithography step and does not require external stressor layers or out-of-plane deflections[22,23,41]. The amount of strain in our structures and the lateral homogeneity of the strain distributions were experimentally determined using Raman spectroscopy, and found to be in good agreement with FEM simulations. Vertical homogeneity of the strain distribution was presumed since the geometry is purely in-plane with no relevant vertical features. Red-shifted, enhanced PL spectra from the strained disks offer further confirmation of the high strain levels in these Ge structures, validating the Raman spectroscopy results. Additionally, the permanent stiction between the Ge nanomembrane and the Si substrate provides excellent thermal conductivity and eliminates heating problems which have traditionally plagued other membrane approaches[20].

The strained Ge structures presented herein also offer an extraordinary level of design flexibility. The size of the structures can also be scaled up or scaled down almost arbitrarily, since the strain is determined almost exclusively by the ratio of the inner microdisk diameter to the total structure diameter. This is important for enabling versatile resonator design if the structures are further patterned into micro-disk or micro-gear resonators, a task which we leave for a future work. Most importantly, the strain can be customized lithographically by changing the dimensions of the structure. This means that multiple strains, and thus multiple bandgaps, can be realized across a single die using a simple one mask process. The presence of multiple bandgaps means that a much wider range of wavelengths can be accessed for emission, modulation and detection, thus raising the possibility of employing these structures in extended wavelength-division multiplexing systems for on-chip optical interconnects. This functionality may find applications not only using Ge but also III-V materials, and our approach should be highly transferable to any arbitrary material system provided that the active material begins with some initial tensile stress.

A particularly critical application is the possibility of using these biaxial strain structures to achieve an efficient, low-threshold Ge laser. Our theoretical investigation on this subject shows that the 1.11% strain which we have observed would increase the fraction of electrons in the direct conduction valley by ~15x compared to the 0.2% strain typically found in epitaxial Ge-on-Si. Moreover, our theoretical modeling shows that this level of strain could reduce the threshold of the demonstrated Ge laser from ~300 kA/cm$^2$ to <1.5 kA/cm$^2$, thereby completing the missing link in a CMOS-compatible on-chip optical interconnect system.

**ASSOCIATED CONTENTS**

**Supporting Information**

Detailed descriptions of fabrication, physical simulation, measurement, theoretical modeling, and additional figures.

**AUTHOR INFORMATION**

**Corresponding authors**

*E-mail: dsukhdeo@stanford.edu

**Author Contributions**

#These authors contributed equally to this work.

**Notes**

The authors declare no competing financial interest.

**ACKNOWLEDGEMENTS**

This work was supported by the Office of Naval Research (grant N00421-03-9-0002) through APIC Corporation (Dr. Raj Dutt) and by a Stanford Graduate Fellowship. The authors would like to thank Dr. Ze Yuan of Stanford University for his assistance implementing in the tight-binding formalism used in our theoretical model of the Ge laser. The authors also thank Dr. Birendra (Raj) Dutt of APIC Corporation and PhotonIC Corporation for his guidance in developing our theoretical model of the Ge laser.

**REFERENCES**

(1) Koo, K.-H.; Cho, H.; Kapur, P.; Saraswat, K. C. *IEEE Trans. Electron Devices* **2007**, *54*, 3206–3215.

(2) Miller, D. A. B. *Proc. IEEE* **2000**, *88*, 728–749.

(3) Soref, R. *Nat. Photonics* **2010**, *4*, 495–497.

(4)   Ishikawa, Y.; Wada, K. *Thin Solid Films* **2010**, *518*, S83–S87.

(5)   Boucaud, P.; Kurdi, M. El; Ghrib, A.; Prost, M.; Kersauson, M. de; Sauvage, S.; Aniel, F.; Checoury, X.; Beaudoin, G.; Largeau, L.; Sagnes, I.; Ndong, G.; Chaigneau, M.; Ossikovski, R. *Photonics Res.* **2013**, *1*, 102–109.

(6)   Dutt, B.; Lin, H.; Sukhdeo, D.; Vulovic, B.; Gupta, S.; Nam, D.; Saraswat, K. C.; Harris, J. S. *IEEE J. Sel. Top. Quantum Electron.* **2013**, *19*, 1502706.

(7)   Nayfeh, A.; Chui, C. O.; Saraswat, K. C.; Yonehara, T. *Appl. Phys. Lett.* **2004**, *85*, 2815.

(8)   Liu, J.; Camacho-Aguilera, R.; Bessette, J. T.; Sun, X.; Wang, X.; Cai, Y.; Kimerling, L. C.; Michel, J. *Thin Solid Films* **2012**, *520*, 3354–3360.

(9)   Van der Walle, C. G. *Phys. Rev. B Condens. matter* **1989**, *39*, 1871–1883.

(10)  Wang, J.; Lee, S. *Sensors* **2011**, *11*, 696–718.

(11)  Vivien, L.; Osmond, J.; Fédéli, J.-M.; Marris-Morini, D.; Crozat, P.; Damlencourt, J.-F.; Cassan, E.; Lecunff, Y.; Laval, S. *Opt. Express* **2009**, *17*, 6252–6257.

(12)  Michel, J.; Liu, J.; Kimerling, L. C. *Nat. Photonics* **2010**, *4*, 527–534.

(13)  Roth, J. E.; Fidaner, O.; Schaevitz, R. K.; Kuo, Y.-H.; Kamins, T. I.; Harris, J. S.; Miller, D. A. B. *Opt. Express* **2007**, *15*, 5851–5859.

(14)  Camacho-Aguilera, R. E.; Cai, Y.; Patel, N.; Bessette, J. T.; Romagnoli, M.; Kimerling, L. C.; Michel, J. *Opt. Express* **2012**, *20*, 11316–11320.

(15)  Liu, J.; Kimerling, L. C.; Michel, J. *Semicond. Sci. Technol.* **2012**, *27*, 094006.

(16)  Liu, J.; Sun, X.; Pan, D.; Wang, X.; Kimerling, L. C.; Koch, T. L.; Michel, J. *Opt. Express* **2007**, *15*, 11272–11277.

(17)  Dutt, B.; Sukhdeo, D. S.; Nam, D.; Vulovic, B. M.; Saraswat, K. C. *IEEE Photonics J.* **2012**, *4*, 2002–2009.

(18)  Greil, J.; Lugstein, a; Zeiner, C.; Strasser, G.; Bertagnolli, E. *Nano Lett.* **2012**, *12*, 6230–6234.

(19)  Süess, M. J.; Geiger, R.; Minamisawa, R. A.; Schiefler, G.; Frigerio, J.; Chrastina, D.; Isella, G.; Spolenak, R.; Faist, J.; Sigg, H. *Nat. Photonics* **2013**, *7*, 466–472.

(20)  Sukhdeo, D. S.; Nam, D.; Kang, J.-H.; Brongersma, M. L.; Saraswat, K. C. *Photonics Res.* **2014**, *2*, A8.

(21)  Huo, Y.; Lin, H.; Chen, R.; Makarova, M.; Rong, Y.; Li, M.; Kamins, T. I.; Vuckovic, J.; Harris, J. S. *Appl. Phys. Lett.* **2011**, *98*, 011111.

(22) Sánchez-Pérez, J. R.; Boztug, C.; Chen, F.; Sudradjat, F. F.; Paskiewicz, D. M.; Jacobson, R. B.; Lagally, M. G.; Paiella, R. *Proc. Natl. Acad. Sci. U. S. A.* **2011**, *108*, 18893–18898.

(23) Nam, D.; Sukhdeo, D.; Roy, A.; Balram, K.; Cheng, S.-L.; Huang, K. C.-Y.; Yuan, Z.; Brongersma, M.; Nishi, Y.; Miller, D.; Saraswat, K. *Opt. Express* **2011**, *19*, 25866–25872.

(24) Capellini, G.; Reich, C.; Guha, S.; Yamamoto, Y.; Lisker, M.; Virgilio, M.; Ghrib, A.; El Kurdi, M.; Boucaud, P.; Tillack, B.; Schroeder, T. *Opt. Express* **2014**, *22*, 399–410.

(25) Minamisawa, R. A.; Süess, M. J.; Spolenak, R.; Faist, J.; David, C.; Gobrecht, J.; Bourdelle, K. K.; Sigg, H. *Nat. Commun.* **2012**, *3*, 1096.

(26) Nam, D.; Sukhdeo, D. S.; Kang, J.-H.; Petykiewicz, J.; Lee, J. H.; Jung, W. S.; Vučković, J.; Brongersma, M. L.; Saraswat, K. C. *Nano Lett.* **2013**, *13*, 3118–3123.

(27) Nam, D.; Sukhdeo, D.; Gupta, S. *IEEE J. Sel. Top. Quantum Electron.* **2013**, *20*.

(28) Okyay, A. K.; Nayfeh, A. M.; Saraswat, K. C.; Yonehara, T.; Marshall, A.; McIntyre, P. C. *Opt. Lett.* **2006**, *31*, 2565–2567.

(29) Perez, R.; Gumbsch, P. *Phys. Rev. Lett.* **2000**, *84*, 5347–5350.

(30) Capellini, G.; Kozlowski, G.; Yamamoto, Y.; Lisker, M.; Wenger, C.; Niu, G.; Zaumseil, P.; Tillack, B.; Ghrib, a.; de Kersauson, M.; El Kurdi, M.; Boucaud, P.; Schroeder, T. *J. Appl. Phys.* **2013**, *113*, 013513.

(31) Nam, D.; Sukhdeo, D.; Cheng, S.-L.; Roy, A.; Chih-Yao Huang, K.; Brongersma, M.; Nishi, Y.; Saraswat, K. *Appl. Phys. Lett.* **2012**, *100*, 131112.

(32) Nataraj, L.; Xu, F.; Cloutier, S. G. *Opt. Express* **2010**, *18*, 7085–7091.

(33) Nam, D.; Sukhdeo, D. S.; Gupta, S.; Kang, J.; Brongersma, M. L.; Saraswat, K. C. *IEEE J. Sel. Top. Quantum Electron.* **2014**, *20*, 1500107.

(34) Boztug, C.; Sánchez-Pérez, J. R.; Sudradjat, F. F.; Jacobson, R. B.; Paskiewicz, D. M.; Lagally, M. G.; Paiella, R. *Small* **2013**, *9*, 622–630.

(35) Jain, J. R.; Ly-Gagnon, D.-S.; Balram, K. C.; White, J. S.; Brongersma, M. L.; Miller, D. A. B.; Howe, R. T. *Opt. Mater. Express* **2011**, *1*, 1121–1126.

(36) Liu, J.; Sun, X.; Camacho-Aguilera, R.; Kimerling, L. C.; Michel, J. *Opt. Lett.* **2010**, *35*, 679–681.

(37) Chang, G.-E.; Chang, S.-W.; Chuang, S. L. *Opt. Express* **2009**, *17*, 11246–11258.

(38) El Kurdi, M.; Fishman, G.; Sauvage, S.; Boucaud, P. *J. Appl. Phys.* **2010**, *107*, 013710.

(39) Pizzi, G.; Virgilio, M.; Grosso, G. *Nanotechnology* **2010**, *21*, 055202.

(40) Claussen, S. A.; Tasyurek, E.; Roth, J. E.; Miller, D. a B. *Opt. Express* **2010**, *18*, 25596–25607.

(41)  El Kurdi, M.; Bertin, H.; Martincic, E.; De Kersauson, M.; Fishman, G.; Sauvage, S.; Bosseboeuf, A.; Boucaud, P. *Appl. Phys. Lett.* **2010**, *96*, 041909.

# Supporting Information

A Nanomembrane-Based Bandgap-Tunable Germanium Microdisk Using Lithographically-Customizable Biaxial Strain for Silicon-Compatible Optoelectronics

*David S. Sukhdeo[†,\*,#], Donguk Nam[†,#], Ju-Hyung Kang[‡],*

*Mark L. Brongersma[‡] and Krishna C. Saraswat[†]*

[†]Department of Electrical Engineering, Stanford University, Stanford, CA 94305, USA,

[‡]Department of Materials Science and Engineering, Stanford University, Stanford, CA 94305, USA

[#]These authors contributed equally to this work

[\*]Email: dsukhdeo@stanford.edu

Contents in this supplement are presented to further strengthen the arguments in the main text.

1. Methods

2. Fabrication process flow (Figure S1)

3. Effect of additional etch slits on the strain distribution (Figure S2)

4. Effect of changing the number of etch slits (Figure S3)

5. Micrographs of fractured microdisk structures (Figure S4)

6. Interaction between strain and doping on the direct conduction valley occupancy

7. References

**1. Methods**

**Device fabrication.** The fabrication process, illustrated in Figure S1, began with a germanium-on-insulator (GOI) substrate consisting of a Si/SiO$_2$/Ge material stack prepared according to the method of Ref 1. This process results in a residual ~0.2% biaxial strain in the Ge nanomembrane immediately after GOI substrate fabrication[1]. A radially symmetric pattern of slits, illustrated in Figure 2, was then dry etched through the Ge using e-beam lithography for patterning. The etch process itself consisted of SF$_6$ and CHClF$_2$ (Freon-22) plasma, with ~300 nm ZEP e-beam resist used as an etch mask. As a final step, the sacrificial oxide (SiO$_2$ in the material stack) was isotropically etched from underneath the entire structure using a hydrofluoric acid (HF) wet etch. The use of a wet HF process results in the Ge being brought into contact with and permanently adhered to the underlying Si substrate due to stiction as the HF evaporates[2].

**Physical simulations.** All FEM simulations of strain distributions in Ge were performed using commercial software, namely COMSOL Multiphysics version 4.4. The "fine" mesh size in COMSOL was used in all FEM simulations. The Ge was assumed to have a 0.2% initial strain, and Ge's Young's Modulus and Poisson ratio were taken to be 103 GPa and 0.26, respectively.

**Optical characterization.** Raman spectroscopy and photoluminescence (PL) measurements were performed using excitation lasers with 514 nm and 532 nm wavelengths, respectively. For both measurements, signals from the samples were collected using a 100x magnification lens and were sent to a spectrometer system. The excitation powers for Raman and PL measurements were <1 mW and 12 mW, respectively. Sample heating is known to be negligible with these excitation powers due to stiction between the structures and the underlying substrate[3].

**Threshold current modeling.** The threshold of a Ge laser was modeled using the same approach that we previously used in Ref 4. However, in this work our $sp^3d^5s^*$ tight-binding model was modified so as to force the band edges of strained Ge to match the values given by the deformation potentials of Ref 5, including a crossover of the direct bandgap at 1.7% biaxial strain. After obtaining the band structures from tight-binding, the relationship between carrier densities and quasi-Fermi levels was obtained by numerical integrations over the entire 1st Brillouin zone. The optical gain was modeled using the empirical absorption coefficient for Ge given in Ref 6 along with the free carrier absorption relation of Ref 7 which is an empirical fit to experimental data. Lasing was assumed to always occur at the wavelength of peak net gain.

Finally, steady-state carrier concentrations were converted into a threshold current density using the recombination coefficients for Ge given in Ref 7.

## 2. Fabrication process flow

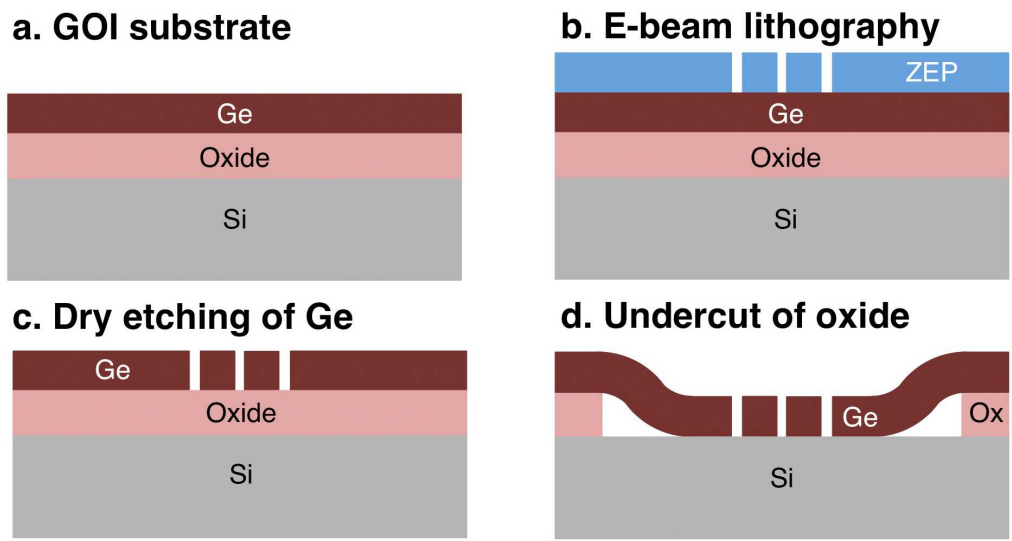

**Figure S1**. Process flow schematic. (a) The initial GOI substrate. (b) E-beam lithography for resist patterning. (c) Dry etching of the Ge layer. (d) Undercut of the sacrificial oxide layer.

## 3. Effect of additional etch slits on the strain distribution

In order to ascertain the impact of the "additional" etch slits on the strain distribution, we have performed two FEM simulations. Both of these FEM simulations are for a 5 μm inner microdisk diameter and a 100 μm total structure diameter, with 20 μm between the total structure dimensions and the fixed boundary. This 20 μm approximates the outward etching of the sacrificial oxide during the final fabrication step. All etch slits were taken to be 500 nm wide with rounded tips on each end, in accordance with our experimental design parameters. In the first FEM simulation, shown in Figure S2a, only the eight main etch slits have been included. In the second FEM simulation, shown in Figure S2b, additional etch slits have been included as they were in our actual devices. It can be readily seen that the stress distribution remains qualitatively extremely similar. Looking at the zoomed-in views of the microdisk region, we find that the simulation without additional etch slits resulted in a 0.590% biaxial strain at the microdisk center, whereas the simulation with the additional etch slits resulted in a 0.582% biaxial strain at the microdisk center. In both cases the maximum biaxial strain in the corner regions was approximately ~0.88%. Thus, we conclude that the only substantive effect of adding additional etch slits is to very slightly reduce the strain at the microdisk center without otherwise altering the overall strain distribution.

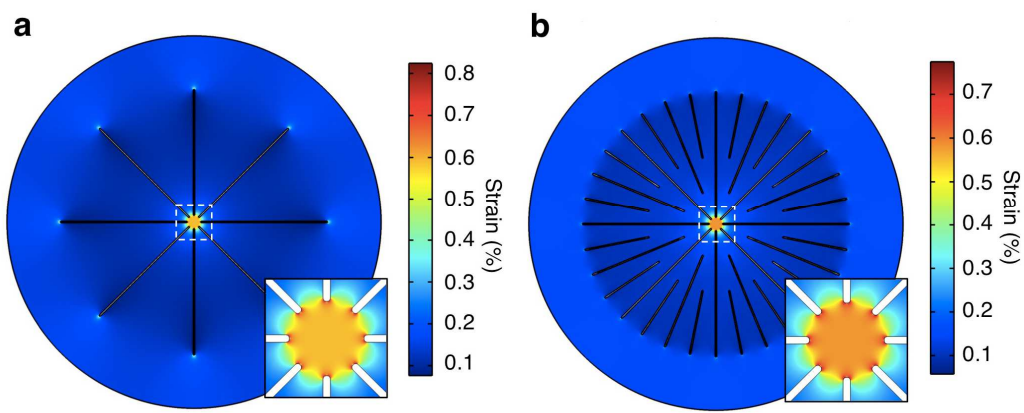

**Figure S2.** Impact of the additional etch slits. (a,b) Finite element method (FEM) simulations of a 100 μm diameter structure with a 5 μm diameter inner microdisk (a) without additional etch slits, and (b) with additional etch slits.

## 4. Effect of changing the number of etch slits

Although our fabricated devices all had exactly eight "main" etch slits touching the inner microdisk, we have performed FEM simulations to investigate how the stress distribution would be affected by changing the number of etch slits which touch the inner microdisk. For these simulations, shown in Figure S3, we have fixed the inner microdisk diameter to 5 μm, the total structure diameter to 50 μm, and assumed a constant 20 μm between the total structure dimensions and the fixed boundary to approximate the outward etching of the sacrificial oxide. All etch slits were again taken to be 500 nm wide with rounded tips on each end, in accordance with our experimental design parameters. Meanwhile, the number of main etch slits was varied from 3 to 20. It is immediately clear from Figure S3a that increasing the number of main etch slits drastically increases the spatial homogeneity of the strain in the microdisk. The relationship between the number of etch slits and the microdisk strain and the corner strain, shown in Figure S3b is more complex. For the design parameters used in this series of FEM simulations it would appear that somewhere between 10 and 15 main etch slits offers the optimal combination of relatively high microdisk strain and relatively low corner strain, though we expect this number to change significantly if the microdisk diameter is changed and/or the width of the etch slits is changed. Nevertheless, this suggests that changing the number of main etch slits may be another avenue to further improve this structure in a future work.

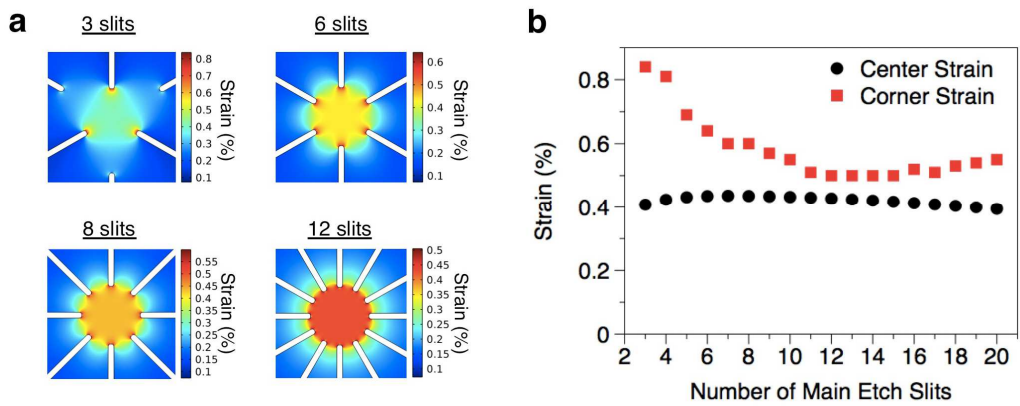

**Figure S3.** Changing the number of main etch slits. (a) Finite element method (FEM) simulations of the structure with different numbers of main etch slits. (b) Biaxial strains at the microdisk center and in the corner regions as a function of the number of main etch slits.

## 5. Micrographs of fractured microdisk structures

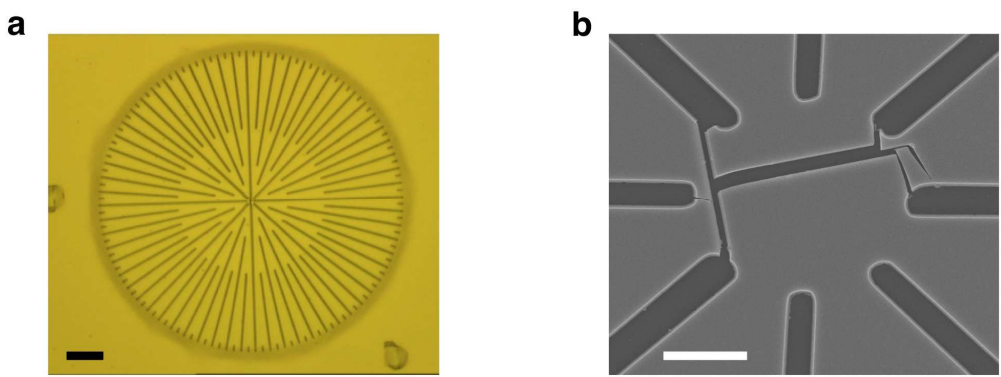

**Figure S4.** Micrographs of fractured microdisk structures. (a) Optical micrograph. Scale bar, 20 μm. (b) Scanning electron micrograph. Scale bar, 2 μm.

**6. Interaction between strain and doping on the direct conduction valley occupancy**

In the main text (Figure 6a) we have shown how the fraction of electrons in the direct conduction valley is enhanced with strain, but that this enhancement depends on the n-type doping level. Although this work does not focus on doping, in the interest of placing this work in context it is instructive to examine how strain and n-type doping interact with each other given n-type doping's pre-eminent role in state of the art Ge photonics.

Since degenerate n-type doping enhances spillover from lower energy conduction valleys to higher energy valleys, once a direct gap is achieved this means that degenerate n-type doping will cause spillover of electrons out of the lower energy direct valley and into the higher energy indirect valleys. This is in addition to the fact that the indirect valleys have a larger density of states and results in a negative interaction between strain and doping which is clearly evidenced in Figure 6a. Specifically, while strain always enhances the fraction of electrons in the direct conduction valley, the relative enhancement diminishes with increasing doping. Applying 1.0% biaxial tensile strain, for instance, results in a ~20x enhancement for undoped Ge, a ~14x enhancement for $5\times10^{19}$ cm$^{-3}$ n-doped Ge, a ~7x enhancement for $1\times10^{20}$ cm$^{-3}$ n-doped Ge, and an only ~3x enhancement for $2\times10^{20}$ cm$^{-3}$ n-doped Ge. (This also corrects a methodological flaw in Ref 8 where we modeled the enhancement from strain in the "heavy doping" case but simply placed the electron quasi-Fermi level at the direct conduction valley edge, a condition which does not correspond to a constant doping level). Moreover, for strains >1.7% degenerate n-type doping actually decreases the fraction of electrons in the direct conduction valley because these strains turn Ge into a direct bandgap semiconductor. A related negative interaction between strain and doping with regard to the threshold of a Ge laser is also expected and has been discussed in depth theoretically in a previous work[4] which concluded that the optimal scenario for a Ge laser is large tensile strain coupled comparatively low n-type doping.

## 7. References (Supporting Information Only)

(1) Jain, J. R.; Ly-Gagnon, D.-S.; Balram, K. C.; White, J. S.; Brongersma, M. L.; Miller, D. A. B.; Howe, R. T. *Opt. Mater. Express* **2011**, *1*, 1121–1126.

(2) Nam, D.; Sukhdeo, D. S.; Gupta, S.; Kang, J.; Brongersma, M. L.; Saraswat, K. C. *IEEE J. Sel. Top. Quantum Electron.* **2014**, *20*, 1500107.

(3) Sukhdeo, D. S.; Nam, D.; Kang, J.-H.; Brongersma, M. L.; Saraswat, K. C. *Photonics Res.* **2014**, *2*, A8.

(4) Dutt, B.; Sukhdeo, D. S.; Nam, D.; Vulovic, B. M.; Saraswat, K. C. *IEEE Photonics J.* **2012**, *4*, 2002–2009.

(5) Van der Walle, C. G. *Phys. Rev. B Condens. matter* **1989**, *39*, 1871–1883.

(6) Liu, J.; Kimerling, L. C.; Michel, J. *Semicond. Sci. Technol.* **2012**, *27*, 094006.

(7) Liu, J.; Sun, X.; Pan, D.; Wang, X.; Kimerling, L. C.; Koch, T. L.; Michel, J. *Opt. Express* **2007**, *15*, 11272–11277.

(8) Nam, D.; Sukhdeo, D.; Roy, A.; Balram, K.; Cheng, S.-L.; Huang, K. C.-Y.; Yuan, Z.; Brongersma, M.; Nishi, Y.; Miller, D.; Saraswat, K. *Opt. Express* **2011**, *19*, 25866–25872.